\documentclass[]{article}

\usepackage{xspace}
\usepackage{url}
\usepackage{proof}
\usepackage{parskip}

\usepackage[outdir=./]{epstopdf}

\usepackage{tabularx}
\usepackage{booktabs}
\usepackage{multirow}
\usepackage{multicol}
\usepackage{pdflscape}

\usepackage{caption}
\usepackage{subcaption}
\usepackage[rightcaption]{sidecap}
\sidecaptionvpos{figure}{c}
\graphicspath{{./fig/}}
\usepackage{amsmath}
\usepackage{amssymb}
\usepackage{syntax}
\usepackage{stmaryrd}
\usepackage{mathtools}

\usepackage{algorithm}
\usepackage{algpseudocode}
\makeatletter
\algrenewcommand\ALG@beginalgorithmic{\small}
\makeatother
\usepackage{listings}
\lstnewenvironment{Bench}[1][]
{\lstset{ %
		showlines=true,
		xleftmargin=5ex,
		firstnumber=1,
		language=java,                % choose the language of the code
		basicstyle=\scriptsize\ttfamily,       % the size of the fonts that are 
		numbers=left,                   % where to put the line-numbers
		numberstyle=\tiny,      % the size of the fonts that are used for the 
		stepnumber=1,                   % the step between two line-numbers.
		numbersep=3pt,                  % how far the line-numbers are from the 
		showspaces=false,               % show spaces adding particular 
		showstringspaces=false,         % underline spaces within strings
		showtabs=false,                 % show tabs within strings adding 
		frame=topline,                   % adds a frame around the code
		frame=shadowbox
		tabsize=2,                      % sets default tabsize to 2 spaces
		captionpos=b,                   % sets the caption-position to bottom
		breaklines=true,                % sets automatic line breaking
		breakatwhitespace=false,        % sets if automatic breaks should only 
		escapeinside={\%*}{*)},          % if you want to add a comment within 
		mathescape=true,
		columns=flexible,
		morekeywords={},
		escapeinside={(*@}{@*)},
		#1
}}{}

\usepackage{listings}
\usepackage{color}
\definecolor{codegreen}{rgb}{0,0.6,0}
\definecolor{codegray}{rgb}{0.5,0.5,0.5}
\definecolor{codepurple}{rgb}{0.58,0,0.82}
\definecolor{backcolour}{rgb}{0.95,0.95,0.92}
\lstdefinestyle{mystyle}{
	backgroundcolor=\color{backcolour},   
	commentstyle=\color{codegreen},
	keywordstyle=\color{magenta},
	numberstyle=\tiny\color{codegray},
	stringstyle=\color{codepurple},
	basicstyle=\footnotesize,
	breakatwhitespace=false,         
	breaklines=true,                 
	captionpos=b,                    
	keepspaces=true,                 
	numbers=left,                    
	numbersep=5pt,                  
	showspaces=false,                
	showstringspaces=false,
	showtabs=false,                  
	tabsize=2
}
\usepackage{tikz}
\usetikzlibrary{fit}					% fitting shapes to coordinates
\usetikzlibrary{backgrounds}	% drawing the background after the foreground
\usetikzlibrary{positioning}	% specify the relative x and y positions separately 
\usetikzlibrary{arrows}
\usetikzlibrary{shapes.geometric}
\definecolor {processblue}{cmyk}{0.96,0,0,0}
\usetikzlibrary{calc}

\usepackage{enumitem}

\usepackage{xcolor} % <======================================= for green
\usepackage[colorlinks]{hyperref} % <====================== last packge to be called
\newcommand{\pipe}{Pipeline}
\newcommand{\Pipe}{Pipeline}
\newcommand{\actor}{operator}
\newcommand{\Actor}{Operator}
\newcommand{\stype}{structure type\xspace}

\newcommand{\dtype}{data type\xspace}
\newcommand{\pico}{PiCo\xspace}

\newcommand{\pname}[1]{{\textsc {#1}}}
\newcommand{\pnew}{\pname{new}\xspace}
\newcommand{\pto}{\pname{to}\xspace}
\newcommand{\ppair}{\pname{pair}\xspace}
\newcommand{\pmerge}{\pname{merge}\xspace}
\newcommand{\pseq}{\farg\vert\farg}
\newcommand{\psum}{+}

\newcommand{\mylist}{\text{list}}
\newcommand{\stream}{\text{stream}}
\newcommand{\bag}{\text{bag}}

\newcommand{\aname}[1]{{\texttt {#1}}}
\newcommand{\map}{\aname{map}\xspace}
\newcommand{\flatmap}{\aname{flatmap}}
\newcommand{\combine}{\aname{combine}}
\newcommand{\reduce}{\aname{reduce}}
\newcommand{\aggregate}{\aname{fold+reduce}}
\newcommand{\emitter}{\aname{emit}}
\newcommand{\collector}{\aname{collect}}
\newcommand{\fromfile}{\aname{from-file}}
\newcommand{\tofile}{\aname{to-file}}
\newcommand{\fromsocket}{\aname{from-socket}}
\newcommand{\tosocket}{\aname{to-socket}}
\newcommand{\fromtwitter}{\aname{from-twitter}}
\newcommand{\bmap}{\aname{b-\map}}

\newcommand{\bcombine}{\aname{b-\combine}}

\newcommand{\zipp}{\aname{zip-}}
\newcommand{\joinp}{\aname{join-}}
\newcommand{\bflatmapzip}{\zipp\flatmap}
\newcommand{\bmapzip}{\zipp\map}
\newcommand{\bflatmapjoin}{\joinp\flatmap}
\newcommand{\bmapjoin}{\joinp\map}
\newcommand{\baggregatejoin}{\joinp\aggregate}
\newcommand{\amodifier}[1]{\aname{{#1}-}}
\newcommand{\windowing}{\amodifier{w}}
\newcommand{\partitioning}{\amodifier{p}}
\newcommand{\winpar}{\windowing\partitioning}
\newcommand{\core}{\amodifier{core}}
\newcommand{\unary}{\amodifier{unary}}
\newcommand{\binary}{\amodifier{binary}}

\newcommand{\wsize}{|W|}
\newcommand{\wslide}{\delta}

\newcommand{\ctype}[2]{{#1}_{#2}}
\newcommand{\ctypeopt}[2]{{#1}_{#2}^\circ}
\newcommand{\tpairio}[2]{{#1} \rightarrow {#2}}
\newcommand{\stsym}{\sigma}

\newcommand{\stall}{\Sigma}
\newcommand{\void}{\emptyset}
\newcommand{\stringtype}{{\text{String}}}

\newcommand{\ocol}[1]{\overrightarrow{#1}}
\newcommand{\Set}[1]{\left\lbrace{#1}\right\rbrace}
\newcommand{\Seq}[1]{\left[{#1}\right]}
\newcommand{\ST}{\bullet}
\newcommand{\Cons}[2]{#1 :: #2}
\newcommand{\CONCAT}{+\!+}
\newcommand{\farg}{\;}
\newcommand{\card}[1]{\left|{#1}\right|}
\newcommand{\natsym}{{\mathbb N}}
\newcommand{\markeq}[1]{\stackrel{({#1})}{=}}

\newcommand{\pair}[2]{\left({#1},{#2}\right)}
\newcommand{\select}[2]{\sigma_{#2}^{#1}}

\newcommand{\inode}[1]{v_I({#1})}
\newcommand{\onode}[1]{v_O({#1})}

\newcommand{\DF}{Dataflow\xspace}

\newcommand{\BD}{Big Data\xspace}

\usepackage{amsthm}

\newtheorem{mydef}{Definition}

\tikzstyle{implies} = [thin,double,double equal sign distance,-implies]
\tikzstyle{mydotted} = [thick,dotted,shorten >=0.2cm,shorten <=0.2cm]
\tikzstyle{edgeattr} = []
\tikzstyle{optedge} = [dotted]
\tikzstyle{matrixattr} = [row sep=0.5em, column sep=1.5em]
\tikzstyle{classmatrixattr} = [row sep=1.5em, column sep=0.1em]
\tikzstyle{line}=[-]
\tikzstyle{classarrow}=[->, >= latex]
\tikzset{
text height=1.5ex,
text depth=0.25ex,
align=center
classical/.style={thin,double,<->,shorten >=4pt,shorten <=4pt,>=stealth}
}

\tikzstyle{syn-pipe}=[rectangle,
minimum size=1cm,
draw=green!80,
fill=green!20]

\tikzstyle{syn-op}=[rectangle,
minimum height=0.8cm,
minimum width=1cm,
draw=lime!80,
fill=lime!20]

\tikzstyle{class}=[rectangle,
minimum height=0.5cm,
minimum width=1cm,
draw=blue!80,
fill=blue!10]

\tikzstyle{ff-actor}=[circle,
minimum size=1cm,
draw=blue!80,
fill=blue!20]

\tikzstyle{ff-pipe}=[rectangle,
minimum height=1cm,
minimum width=1.5cm,
draw=orange!80,
fill=orange!25]

\tikzstyle{ff-op}=[diamond,
minimum height=1cm,
minimum width=1cm,
draw=yellow!80,
fill=yellow!25]

\tikzstyle{ff-empty-pipe}=[rectangle,
minimum height=1cm,
minimum width=1.5cm]

\tikzstyle{block}=[rectangle,
minimum size=1cm,
draw=black, dotted]

\newcommand{\GT}[1]{\smallskip\noindent{\small\bf [[#1 {\tiny(Guy~T.)}]]}}
\newcommand{\MD}[1]{\smallskip\noindent{\small\bf [[#1 {\tiny(M.~D.)}]]}}
\newcommand{\CM}[1]{\smallskip\noindent{\footnotesize \textcolor{blue}{ [[#1 {\tiny(C.~M.)}]]}}}
\newcommand{\ma}[1]{\smallskip\noindent \textcolor{violet}{{\small\bf [[#1 {\tiny(M.~A.)}]]}}}
\newcommand{\GUY}[1]{{\footnotesize \textcolor{red}{[[(Guy T.) #1]]}}}
\newcommand{\Guy}[1]{{\footnotesize \textcolor{green}{[[(Guy T.) #1]]}}}

\renewcommand{\GT}[1]{}
\renewcommand{\MD}[1]{}
\renewcommand{\CM}[1]{}
\renewcommand{\ma}[1]{}
\renewcommand{\GUY}[1]{}
\renewcommand{\Guy}[1]{}

\title{A Formal Semantics for Data Analytics Pipelines\\[2ex] Technical Report}
\author{Maurizio Drocco, Claudia Misale, Guy Tremblay, Marco Aldinucci}

\begin{document}
\maketitle

\begin{abstract}
In this report, we present a new programming model based on Pipelines and Operators, which are the building blocks of programs written in \pico, a DSL for Data Analytics Pipelines.
In the model we propose, we use the term \pipe{} to denote a workflow that processes data collections---rather than a computational process---as is common in the data processing community.

The novelty with respect to other frameworks is that all \pico operators are polymorphic with respect to data types. 
This makes it possible to 1) re-use the same algorithms and pipelines on different data models (e.g., streams, lists, sets, etc); 2)~reuse the same operators in different contexts, and 3) update operators without affecting the calling context, i.e., the previous and following stages in the pipeline. Notice that in other mainstream frameworks, such as Spark, the update of a pipeline by changing a transformation with another is not necessarily trivial, since it may require the development of an input and output proxy to adapt the new transformation for the calling context.

In the same line, we provide a formal framework (i.e., typing and semantics) that characterizes programs from the perspective of how they transform the data structures they process---rather than the computational processes they represent.
This approach allows to reason about programs at an abstract level, without taking into account any aspect from the underlying execution model or implementation.

\end{abstract}

\section{Introduction}

% introduction to problem statement
Big Data is becoming one of the most (ab)used buzzword of our times. In companies, industries, academia, the interest is dramatically increasing and everyone wants to ``do \BD'', even though its definition or role in analytics is not completely clear.
From a high-level perspective, \BD is about extracting knowledge from both structured and unstructured data.
This is a useful process for big companies such as banks, insurance, telecommunication, public institutions, and so on, as well as for business in general.
Extracting knowledge from \BD requires tools satisfying strong requirements with respect to  programmability --- that is, allowing to easily write programs and algorithms to analyze data --- and performance, ensuring scalability when running analysis and queries on multicore or cluster of multicore nodes. 
Furthermore, they need to cope with input data in different formats, e.g. batch from data marts, live stream from the Internet or very high-frequency sources.
In the last decade, a large number of frameworks for \BD processing has been implemented addressing these issues.

Their common aim is to ensure ease of programming by providing a unique framework addressing both batch and stream processing.
Even if they accomplish this task, they often lack of  a clear semantics of their programming and execution model.
For instance, users can be provided with two different data models for representing collections and streams, both supporting the same operations but often having different semantics.

% introduction to pico
We advocate a new Domain Specific Language (DSL), called \textbf{{Pi}}peline \textbf{{Co}}m\-po\-si\-tion (\pico), designed over the presented layered \DF conceptual framework~\cite{17:bigdatasurvey:PPL}. \pico programming model aims at \emph{easing the programming}
% and \emph{enhancing the performance} 
of Analytics applications by two design routes: 1) unifying data access model, and 2) decoupling processing from data layout.

Both design routes undertake the same goal, which is the raising of the level of abstraction in the  programming and the execution model with respect to mainstream approaches in tools (Spark~\cite{zaharia:resilient:2012}, Storm~\cite{Anis:CoRR:storm:15}, Flink~\cite{flink-web} and Google Dataflow~\cite{Dataflow:Akidau:2015}) for \BD analytics, which typically force the specialization of the algorithm to match the data access and layout. Specifically, data transformation functions (called \emph{operators} in \pico) exhibit a different functional types when accessing data in different ways. 

For this reason, the source code should be revised when switching from one data model to the next. This happens in all the above mentioned frameworks and also in the abstract \BD architectures, such as the Lambda~\cite{15:lambda:kiran} and Kappa architectures~\cite{kappa-web}. 
Some of them, such as the Spark framework, provide the runtime with a module to convert streams into micro-batches (Spark Streaming, a library running on Spark core), but still a different code should be written at user-level. The Kappa architecture advocates the opposite approach, i.e., to ``streamize'' batch processing, but the streamizing proxy has to be coded.  The Lambda architecture  requires the implementation of both a batch-oriented and a stream-oriented algorithm, which means coding and maintaining two codebases per algorithm. 

\pico fully decouples algorithm design from data model and layout. Code is designed in a fully functional style by composing stateless \emph{operators} (i.e., transformations in Spark terminology). As we discuss in this report, all \pico operators are polymorphic with respect to data types. This makes it possible to 1) re-use the same algorithms and pipelines on different data models (e.g., streams, lists, sets, etc); 2)~reuse the same operators in different contexts, and 3) update operators without affecting the calling context, i.e., the previous and following stages in the pipeline. Notice that in other mainstream frameworks, such as Spark, the update of a pipeline by changing a transformation with another is not necessarily trivial, since it may require the development of an input and output proxy to adapt the new transformation for the calling context.

This report proceeds as follows.
We formally define the syntax of a program, which is based on \pipe s and \actor s whereas it hides the data structures produced and generated by the program. %We define the Program Semantics layer of the \DF stack as it has been defined in~\cite{17:bigdatasurvey:PPL}.
Then we provide the formalization of a minimal type system defining legal compositions of \actor s into \pipe s.
Finally, we provide a semantic interpretation that maps any \pico program to a functional \DF graph, representing the transformation flow followed by the processed collections.

%----------------------------------------------------------------------------------------
%	SECTION 1
%----------------------------------------------------------------------------------------

\section{Syntax}
\label{ch:pm:design}

We propose a programming model for processing data collections, based on the \DF model. 
The building blocks of a \pico program are \emph{\Pipe s} and \emph{\Actor s},  which we investigate in this section. Conversely, \emph{Collections} are not included in the syntax and they are introduced in Section~\ref{ch:pm:coll} since they contribute at defining the type system and the semantic interpretation of \pico programs.

\subsection{\Pipe s}
\label{ch:pm:pipe}

\begin{figure}[H]
	\centering
	% source pipe
\begin{subfigure}[b]{0.3\textwidth}
\centering
\begin{tikzpicture}
% "text height" and "text depth" are required to vertically
% align the labels with and without indices.

\node (dummy) {};
\node (e) [syn-op, right=0.5 of dummy] {$op$};
\node (post) [right=0.5 of e] {};

\path[->]
(e) edge[edgeattr] (post);

\begin{pgfonlayer}{background}
\node [syn-pipe, fit=(e)] {};
\end{pgfonlayer}

\end{tikzpicture}
\caption{Source}
\label{fig:pipe-source}
\end{subfigure}
%
%
%
% sink pipe
\begin{subfigure}[b]{0.3\textwidth}
\centering
\begin{tikzpicture}
    % "text height" and "text depth" are required to vertically
    % align the labels with and without indices.

\node (pre) {};
\node (e) [syn-op, right=0.5 of dummy] {$op$};
\node (dummy) [right=0.5 of e] {};

\path[->]
(pre) edge[edgeattr] (e);

\begin{pgfonlayer}{background}
\node [syn-pipe, fit=(e)] {};
\end{pgfonlayer}

\end{tikzpicture}
\caption{Sink}
\label{fig:pipe-sink}
\end{subfigure}
%
%
%
% processing pipe
\begin{subfigure}[b]{0.3\textwidth}
\centering
\begin{tikzpicture}
    % "text height" and "text depth" are required to vertically
    % align the labels with and without indices.
   
\node (pre) [] {};
\node (a) [syn-op, right = 0.5 of pre] {$op$};
\node (post) [right = 0.5 of a] {};

\path[->]
    (pre) edge[edgeattr] (a)
    (a) edge[edgeattr] (post)
;

\begin{pgfonlayer}{background}
\node [syn-pipe, fit=(a)] {};
\end{pgfonlayer}

\end{tikzpicture}
\caption{Processing}
\label{fig:pipe-unary}
\end{subfigure}

\bigskip
% linear pipe
\begin{subfigure}[b]{0.4\textwidth}
\centering
\begin{tikzpicture}
    % "text height" and "text depth" are required to vertically
    % align the labels with and without indices.
 
\matrix[column sep=1.5em] {
\node (pre) {}; &
\node (p) [syn-pipe] {$p$}; &
\node (p1) [syn-pipe] {$p_1$}; &
\node (post) {};\\
};

\path[->]
(pre) edge[edgeattr, optedge] (p)
(p) edge[edgeattr] (p1)
(p1) edge[edgeattr, optedge] (post)
;

\begin{pgfonlayer}{background}
\node [syn-pipe, fit=(p) (p1)] {};
\end{pgfonlayer}

\end{tikzpicture}
\caption{Linear \pto}
\label{fig:pipe-non-linear}
\end{subfigure}
%
%
%
%
%
% non-linear pipe
\begin{subfigure}[b]{0.5\textwidth}
\centering
\begin{tikzpicture}
    % "text height" and "text depth" are required to vertically
    % align the labels with and without indices.
 
\matrix[row sep=0.5em, column sep=1.5em] {
&&&
\node (p1) [syn-pipe]{$p_1$};
&&\\

\node (pre) {}; &
\node (p) [syn-pipe] {$p$}; &
\node (bcast) {$\cdot$}; &
&
\node (merge) {$\cdot$};&
\node (post) {};\\

&&&
\node (p2) [syn-pipe]{$p_n$};
&&\\
};

\path[->]
(pre) edge[edgeattr, optedge] (p)
(p) edge[edgeattr] (bcast)
(bcast) edge[edgeattr] (p1)
(bcast) edge[edgeattr] (p2)
(p1) edge[edgeattr, optedge] (merge)
(p2) edge[edgeattr, optedge] (merge)
(merge) edge[edgeattr, optedge] (post)
;

\draw[mydotted] (p1) -- (p2);

\begin{pgfonlayer}{background}
\node [syn-pipe, fit=(p) (p1) (p2) (merge)] {};
\end{pgfonlayer}

\end{tikzpicture}
\caption{Non-linear \pto}
\label{fig:pipe-linear}
\end{subfigure}

\bigskip
% pair
\begin{subfigure}[b]{0.45\textwidth}
\centering
\begin{tikzpicture}
    % "text height" and "text depth" are required to vertically
    % align the labels with and without indices.
 
\matrix[row sep=0.5em, column sep=1.5em] {
\node (pre1) {};&
\node (p1) [syn-pipe] {$p_1$}; &
&\\

&&
\node (a) [syn-op]{$op$}; &
\node (post) {};
&\\

\node (pre2) {};&
\node (p2) [syn-pipe] {$p_2$}; &
&\\
};

\path[->]
(pre1) edge[edgeattr, optedge] (p1)
(pre2) edge[edgeattr, optedge] (p2)
(p1) edge[edgeattr] (a)
(p2) edge[edgeattr] (a)
(a) edge[edgeattr, optedge] (post)
;

\begin{pgfonlayer}{background}
\node [syn-pipe, fit=(p1) (p2) (a)] {};
\end{pgfonlayer}

\end{tikzpicture}
\caption{\ppair}
\label{fig:pipe-pair}
\end{subfigure}
%
%
%
%
%
% non-linear pipe
\begin{subfigure}[b]{0.45\textwidth}
\centering
\begin{tikzpicture}
    % "text height" and "text depth" are required to vertically
    % align the labels with and without indices.
 
\matrix[row sep=0.5em, column sep=1.5em] {
\node (pre1) {};&
\node (p1) [syn-pipe] {$p_1$}; &
&\\

&&
\node (merge) {$\cdot$}; &
\node (post) {};
&\\

\node (pre2) {};&
\node (p2) [syn-pipe] {$p_2$}; &
&\\
};

\path[->]
(pre1) edge[edgeattr, optedge] (p1)
(pre2) edge[edgeattr, optedge] (p2)
(p1) edge[edgeattr] (merge)
(p2) edge[edgeattr] (merge)
(merge) edge[edgeattr] (post)
;

\begin{pgfonlayer}{background}
\node [syn-pipe, fit=(p1) (p2) (merge)] {};
\end{pgfonlayer}

\end{tikzpicture}
\caption{\pmerge}
\label{fig:pipe-merge}
\end{subfigure}
	\caption{Graphical representation of \pico \pipe s\label{fig:pipes}}
\end{figure}
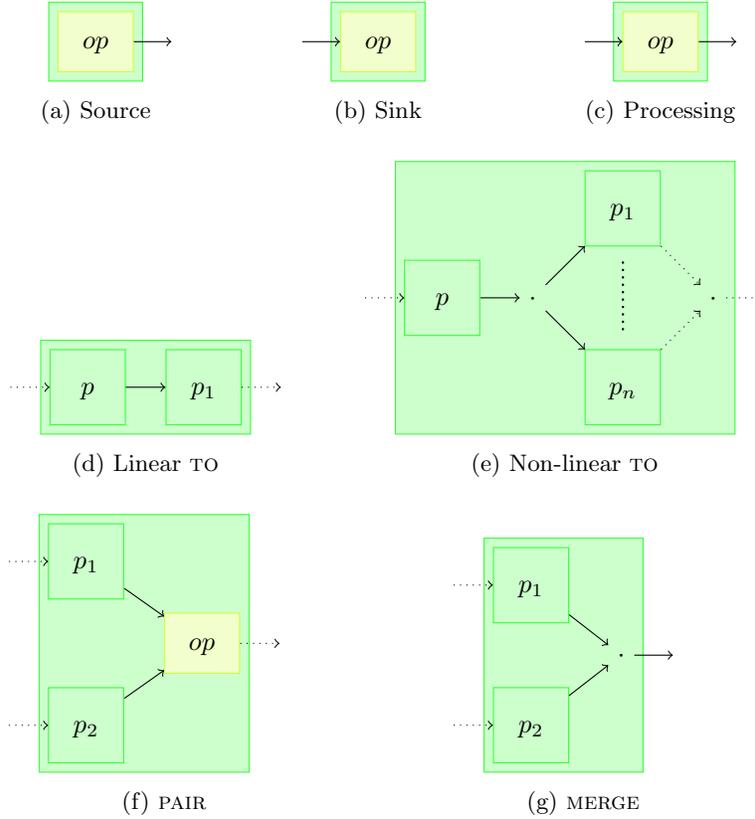

\begin{table}
	\centering
	\footnotesize
	\begin{tabularx}{\textwidth}{lXp{0.3\linewidth}}
		\toprule
		\Pipe & Structural\newline properties & Behavior \\
		
		\midrule
		$\pnew\farg op$ & - &
		data is processed by \actor\ $op$ (i.e., \emph{unary} \pipe{})\\
		
		\midrule
		$\pto\farg p\farg p_1\farg \ldots\farg p_n$
		& associativity for linear \pipe s:\newline
		$
		\begin{array}{l}
		\pto\farg(\pto\farg p_A\farg p_B)\farg p_C\equiv\\
		\pto\farg p_A\farg (\pto\farg p_B\farg p_C)\equiv\\
		p_A \pseq p_B \pseq p_C
		\end{array}
		$
		\smallskip
		\newline
		destination commutativity:\newline
		$
		\begin{array}{l}
		\pto\farg p\farg p_1 \ldots p_n\equiv\\
		\pto\farg p\farg p_{\pi(1)} \ldots p_{\pi(n)}
		\end{array}
		$
		\newline
		for any $\pi$ permutation of $1..n$ &
		data from \pipe\ $p$ is sent to all \pipe s $p_i$ (i.e., broadcast)\\
		
		\midrule
		$\ppair\farg p_1\farg p_2\farg op$	& - &
		data from \pipe s $p_1$ and $p_2$ are pair-wise processed by \actor\ $op$\\
		
		\midrule
		$\pmerge\farg p_1\farg p_2$ & associativity:\newline
		$
		\begin{array}{l}
		\pmerge\farg(\pmerge\farg p_1\farg p_2)\farg p_3\equiv\\
		\pmerge\farg p_1\farg(\pmerge\farg p_2\farg p_3)\equiv\\
		p_1\farg\psum\farg p_2\farg\psum\farg p_3
		\end{array}
		$
		\smallskip
		\newline
		commutativity:\newline
		$
		\begin{array}{l}
		\pmerge\farg p_1\farg p_2\equiv\\
		\pmerge\farg p_2\farg p_1
		\end{array}
		$
		& 	data from \pipe s $p_1$ and $p_2$ are merged, respecting the ordering in case of ordered collections\\
		
		\bottomrule
	\end{tabularx}
	\caption{\Pipe s} 
	\label{tab:pipelines}
\end{table}

The cornerstone concept in the Programming Model is the \emph{\Pipe}, basically a DAG-composition of processing \emph{\actor{s}}.
\Pipe s are built according to the following grammar\footnote{For simplicity, here we introduce the non-terminal $\unary\actor{}$ (resp.\ $\binary\actor$) that includes core and partitioning unary (resp.\ binary) \actor s.}:
\begin{grammar}
	<\pipe> ::= $\pnew\farg<\unary\actor>$
	\alt $\pto\farg<\pipe>\farg<\pipe>\farg\ldots\farg<\pipe>$
	\alt $\ppair\farg<\pipe>\farg<\pipe>\farg<\binary\actor>$
	\alt $\pmerge\farg<\pipe>\farg<\pipe>$
\end{grammar}

We categorize \pipe s according to the number of collections they take as input and output:
\begin{itemize}
	\item A source \pipe{} takes no input and produces one output collection
	\item A sink \pipe{} consumes one input collection and produces no output
	\item A processing \pipe{} consumes one input collection and produces one output collection
\end{itemize}

A pictorial representation of \Pipe s is reported in Figure~\ref{fig:pipes}.
We refer to Figs.~\ref{fig:pipe-source}, \ref{fig:pipe-sink} and \ref{fig:pipe-unary} as \emph{unary} \pipe s, since they are composed by a single \actor.
Figs.~\ref{fig:pipe-linear} and \ref{fig:pipe-non-linear} represent, respectively, linear (i.e., one-to-one) and branching (i.e., one-to-$n$) \pto{} composition.
Figs.~\ref{fig:pipe-pair} and \ref{fig:pipe-merge} represent composition of \pipe s by, respectively, pairing and merging.
A dotted line means the respective path may be void (e.g., a source \pipe{} has void input path).
Moreover, as we show in Section~\ref{ch:pm:type system}, \pipe s are not allowed to consume more than one input collection, thus both \ppair{} and \pmerge{} \pipe s must have at least one void input path.

The meaning of each \pipe{} is summarized in Table~\ref{tab:pipelines}.

\subsection{\Actor{s}}
\label{ch:pm:actors}

\Actor{s} are the building blocks composing a \pipe.
They are categorized according to the following grammar of core \actor\ families:
\begin{grammar}
	<$\core\actor$> ::= $<\core\unary\actor>$ | $<\core\binary\actor>$
	
	<$\core\unary\actor$> ::= $<\map>$ | $<\combine>$ | $<\emitter>$ | $<\collector>$
	
	<\core\binary\actor> ::= $<\bmap>$ | $<\bcombine>$
\end{grammar}

The intuitive meanings of the core \actor s are summarized in Table~\ref{tab:actors}.
\begin{table}[H]
	\centering
	\footnotesize
	\begin{tabularx}{\textwidth}{lp{0.16\linewidth}lX}
		\toprule
		{\Actor\ family} & Categorization & Decomposition & Behavior \\
		
		\midrule
		\map & unary, \newline element-wise & no &
		applies a user function to each element in the input collection\\
		
		\midrule
		\combine & unary, \newline collective& yes &
		synthesizes all the elements in the input collection
		into an atomic value, according to a user-defined policy\\
		
		\midrule
		\bmap & binary, \newline pair-wise & yes &
		the binary counterpart of \map: applies a (binary) user function to 
		each pair generated by pairing (i.e. zipping/joining) two input 
		collections\\
		
		\midrule
		\bcombine & binary, \newline collective & yes &
		the binary counterpart of \combine: synthesizes all pairs generated by pairing (i.e. zipping/joining) two input collections \\
		
		\midrule
		\emitter & produce-only & no &
		reads data from a source, e.g., regular collection, text file, 
		tweet feed, etc.\\
		
		\midrule
		\collector & consume-only & no &
		writes data to some destination, e.g., regular collection, text 
		file, screen, etc.\\
		\bottomrule
	\end{tabularx}
	\caption{Core \actor\ families.\label{tab:actors}}
\end{table}

In addition to core \actor s, generalized \actor s can decompose their input collections by:
\begin{itemize}
	\item partitioning the input collection according to a user-defined 
	grouping 
	policy (e.g., group by key)
	\item windowing the \emph{ordered} input collection according to a 
	user-defined 
	windowing policy (e.g., sliding windows)
\end{itemize}
The complete grammar of \actor s follows:
\begin{grammar}
	<\actor> ::= $<\core\actor>$
	\alt $<\windowing\actor>$ | $<\partitioning\actor>$ | $<\winpar\actor>$
\end{grammar}
where \windowing{} and \partitioning{} denote decomposition by windowing and partitioning, respectively.

For those \actor s $op$ not supporting decomposition (cf.\ 
Table~\ref{tab:actors}), the following structural equivalence holds:
$op\equiv \windowing op \equiv \partitioning op \equiv \windowing\partitioning op$.

\subsubsection{Data-Parallel Operators}
\MD{TODO: establish the analogy between \map/\flatmap\
	and the scalar component of list homomorphisms as defined
	by Gorlatch et al. in HLPP 2016.}

\Actor s in the \map family  are defined according to the 
following 
grammar:
\begin{grammar}
	<\map> ::= $\map\farg f$ | $\flatmap\farg f$
\end{grammar}
where $f$ is a user-defined function (i.e., the \emph{kernel} function) from a host language.\footnote{Note that we treat kernels as terminal symbols, thus we do not define the language in which kernel functions are defined; we rather denote this aspect to a specific implementation of the model.}
The former produces exactly one output element from each input element 
(one-to-one user function), whereas the latter produces a (possibly empty) bounded sequence
of 
output elements for each input element (one-to-many user function) and the 
output collection is the merging of the output sequences.

\Actor s in the \combine\ family synthesize all the elements from an input collection into a single value, according to a user-defined kernel.
They are defined according to the following grammar:
\begin{grammar}
	<\combine> ::= $\reduce\farg \oplus$
	| $\aggregate\farg \oplus_1\farg z\farg \oplus_2$
\end{grammar}
The former corresponds to the classical reduction, whereas the latter is a 
two-phase aggregation that consists in the reduction of partial accumulative states (i.e., partitioned folding with explicit initial value).
The parameters for the \aggregate\ \actor\ specify the initial value for each partial accumulator ($z \in S$, the initial value for the folding), how each input item affects the aggregative state ($\oplus_1:S\times T \to S$, the folding function) and how aggregative states are combined into a final accumulator ($\oplus_2:S\times S \to S$, the reduce function).

\subsubsection{Pairing}
\Actor{s} in the \bmap\ family are intended to be the binary 
counterparts of \map\ \actor s:
\begin{grammar}
	<\bmap> ::= $\bmapzip\farg f$ | $\bmapjoin\farg f$
	\alt $\bflatmapzip\farg f$ | $\bflatmapjoin\farg f$
\end{grammar}
The binary user function $f$ takes as input pairs of elements, one from each of the input collections.
Variants $\zipp$ and $\joinp$ corresponds to the following pairing policies, respectively:
\begin{itemize}
	\item zipping of ordered collections produces the pairs of elements with the same position within the order of respective collections
	\item joining of bounded collections produces the Cartesian product of the input collections
\end{itemize}

Analogously, \actor s in the \bcombine\ family are the binary counterparts of \combine\ \actor s.

\subsubsection{Sources and Sinks}
\Actor s in the \emitter{} and \collector{} families model data collection sources and sinks, respectively:
\begin{grammar}
	<\emitter> ::= $\fromfile\farg$file | $\fromsocket\farg$socket | $\ldots$
	
	<\collector> ::= $\tofile\farg$file | $\tosocket\farg$socket | $\ldots$
\end{grammar}

\subsubsection{Windowing}
\label{ch:pm:windowing}
Windowing is a well-known approach for overcoming the difficulties stemming from the unbounded nature of stream processing.
The basic idea is to process parts of some recent
stream history upon the arrival of new stream items, rather than store
and process the whole stream each time.

\GT{Subsequent means (I think?) ``that come afterward'', ``later''. I think you
	probably mean ``consecutive''?}

\GT{But still\ldots\ not sure of the above sentence.  How about simply
	the following: The basic idea is to process parts of some recent
	stream history upon the arrival of new stream items, rather than store
	and process the whole stream each time.}

\MD{Much better!}

A windowing \actor{} takes an ordered collection, produces a collection (with the same \stype as the input one) of windows (i.e., lists), and applies the subsequent operation to each window.
Windowing \actor s are defined according to the following grammar, where $\omega$ is the windowing policy:
\begin{grammar}
	<$\windowing\actor$> ::= $\windowing<\core\actor>\farg\omega$
\end{grammar}

Among the various definitions from the literature, for the sake of simplicity we only consider policies producing \emph{sliding windows}, characterized by two parameters, namely, a window size $\wsize$---specifying which elements fall into a window---and a sliding factor $\wslide$---specifying how the window slides over the stream items.
Both parameters can be expressed either in time units (i.e., time-based windowing) or in number of items (i.e., count-based windowing).
In this setting, a windowing policy $\omega$ is a term $(\wsize,\wslide,b)$ where $b$ is either {\tt time} or {\tt count}.
A typical case is when $\wsize = \wslide$, referred as a \emph{tumbling} policy.

The meaning of the supported windowing policies will be detailed in semantic terms (Section~\ref{ch:pm:semantic collections}).
Although the \pico syntax only supports a limited class of windowing policies, the semantics we provide is general enough to express other policies such as session windows~\cite{googlecloud:2015}.

As we will show in Section~\ref{ch:pm:type system}, we rely on tumbling windowing to extend bounded \actor s\footnote{We say an \actor\ is \emph{bounded} if it can only deal with bounded collections.} and have them deal with unbounded collections; for instance, \combine{} \actor s are bounded and require windowing to extend them to unbounded collections.

\subsubsection{Partitioning}
\label{ch:pm:partitioning}
Logically, partitioning \actor s take a collection, produces a set (one per 
group) of sub-collections (with the same type as the input one) and applies
the subsequent operation to each sub-collection.
Partitioning \actor s are defined according to the following grammar, where $\pi$ is a user-defined partitioning policy that maps each item to the respective sub-collection:
\begin{grammar}
	<$\partitioning\actor$> ::= $\partitioning<\core\actor>\farg \pi$
\end{grammar}

\Actor s in the \combine, \bmap{} and \bcombine{} families support partitioning, so, for instance, a 
\partitioning\combine\ produces a \bag{} of values, each being the synthesis 
of one group; also the natural join operator from the relational algebra is a particular case of per-group joining.

The decomposition by both partitioning and windowing considers the former 
as the external decomposition, thus it logically produces a set (one per 
group) of collections of windows:
\begin{grammar}
	<$\winpar\actor$> ::= $\winpar<\core\actor>\farg\pi\farg\omega$
\end{grammar}

\subsection{Running Example: The \pname{word-count} \pipe}
\begin{algorithm}
	\caption{A \pname{word-count} \pipe}
	\label{alg:wc model}
	\begin{algorithmic}[0]
		\State $f = \lambda l.\text{list-map}\farg(\lambda w.\pair w 
		1)\farg(\text{split}\farg l)$
		\State $\aname{tokenize} = \flatmap\farg f$
		\newline
		\State $\oplus = \lambda x y.\pair {\pi_1(x)} {\pi_2(x) + 
			\pi_2(y)}$
		\State $\aname{keyed-sum} = \partitioning(\reduce\farg\oplus)\farg\pi_1$
		\newline
		\State $\aname{file-read} = \fromfile\farg$input-file
		\State $\aname{file-write} = \tofile\farg$output-file
		\newline
		\State $\pname{word-count} = \pnew\farg\aname{tokenize} \pseq 
		\pnew\farg\aname{keyed-sum}$
		\State $\pname{file-word-count} = \pnew\ \aname{file-read} \pseq 
		\pname{word-count} \pseq \pnew\farg\aname{file-write}$
	\end{algorithmic}
\end{algorithm}

\GUY{Below: I suggest ``in a generic functional language''
	instead. Because saying it is like lambda-calculus may not be right,
	and may be interpreted rather strictly by some.}
\MD{I agree.}

We illustrate a simple \pname{word-count} \pipe{} in Algorithm~\ref{alg:wc model}.
We assume an hypothetical \pico implementation where the host language provides some common functions over basic types---such as strings and lists---and a syntax for defining and naming functional transformations.
In this setting, the functions $f$ and $\oplus$ in the example are user-defined kernels (i.e., functional transformations) and:
\begin{itemize}
	\item split is a host function mapping a text line (i.e., a string) into the list of words occurring in the line
	
	\item list-map is a classical host map over lists
	\item $\pi_1$ is the left-projection partitioning policy (cf.\ example below, Section~\ref{ch:pm:semantic collections}, Definition~\ref{def:selection})
\end{itemize}

The \actor s have the following meaning:
\begin{itemize}
	\item \aname{tokenize} is a \flatmap\ \actor\ that receives lines $l$ of text and produces, for each word $w$ in each line, a pair $\pair w 1$;
	\item \aname{keyed-sum} is a \partitioning\reduce\ \actor\ that partitions 
	the 
	pairs based on $w$ (obtained with $\pi_1$, using group-by-word) and then sums up each group to 
	$\pair 
	w {n_w}$, where $w$ occurs $n_w$ times in the input text;
	\item \aname{file-read} is an \emitter\ \actor\ that reads from a text file and generates a list of lines;
	
	\item \aname{file-write} is a \collector\ \actor\ that writes a bag of 
	pairs 
	$\pair w {n_w}$ to a  text file.
\end{itemize}

%----------------------------------------------------------------------------------------
%	SECTION 2
%----------------------------------------------------------------------------------------

\section{Type System}
\label{ch:pm:type system}
Legal \pipe{s} are defined according to typing rules, described below.
We denote the typing relation as $a:\tau$, if and only if there exists a legal inference assigning type $\tau$ to the term $a$.

\subsection{Collection Types}
\label{ch:pm:coll}
We mentioned earlier (Section~\ref{ch:pm:design}) that collections are \emph{implicit} entities that flow across \pipe{s} through the DAG edges.
A collection is either \emph{bounded} or \emph{unbounded}; moreover, it is also either \emph{ordered} or \emph{unordered}.
A combination of the mentioned characteristics defines the \emph{\stype} of a collection.
We refer to each \stype with a mnemonic name:
\begin{itemize}
	\item a bounded, ordered collection is a \emph{\mylist}
	\item a bounded, unordered collection is a (bounded) \emph{\bag}
	\item an unbounded, ordered collection is a \emph{\stream}
	%	\item an unbounded, unordered collection is an \emph{unbounded \bag}
\end{itemize}
%\ma{the 4th option appears impossible, explain in few words why not}

%\GUY{I agree. This ties with my remarks a while ago about why all
%types had a specific name\ldots\ except this one type}

A collection type is characterized by its \stype and its \emph{\dtype}, namely the type of the collection elements.
Formally, a collection type has form $\ctype{T}{\stsym}$ where 
$\stsym\in\stall$ is the \stype, $T$ is the \dtype---and where $\stall=\{\bag,\mylist,\stream\}$ is the set of all \stype{s}.
We also partition $\stall$ into $\stall_b$ and $\stall_u$, defined as the 
sets of bounded and unbounded \stype{s}, respectively.
Moreover, we define $\stall_o$ as the set of ordered \stype{s}, thus 
$\stall_b \cap \stall_o = \{\mylist\}$ and $\stall_u \cap \stall_o = \{\stream\}$.
Finally, we allow the void type $\void$.

\subsection{\Actor{} Types}
\begin{table}
	\centering
	\footnotesize
	\begin{tabular}{lc}
		\toprule
		\Actor & Type \\
		
		\midrule
		\multicolumn{2}{c}{Unary}\\
		\midrule
		$\map$ & $\tpairio {\ctype T \stsym} {\ctype U \stsym}, \forall\stsym 
		\in 
		\stall$\\
		
		$\combine$, $\partitioning\combine$ & $\tpairio {\ctype T \stsym} 
		{\ctype U 
			\stsym}, \forall\stsym \in \stall_b$\\
		
		$\windowing\combine$, $\winpar\combine$ & $\tpairio 
		{\ctype 
			T {\stsym}} {\ctype U {\stsym}}, \forall\stsym \in \stall_o$\\
		
		$\emitter$ & $\tpairio \void {\ctype U \stsym}$ \\
		$\collector$ & $\tpairio {\ctype T \stsym} \void$ \\
		
		\midrule
		\multicolumn{2}{c}{Binary}\\
		\midrule
		$\bmap$, $\partitioning\bmap$ & $\tpairio {\ctype {T} {\stsym} \times 
			\ctype {T'} {\stsym}} {\ctype U {\stsym}}, 
		\forall\stsym\in\stall_b$ \\
		
		$\windowing\bmap$, $\winpar\bmap$ & $\tpairio {\ctype 
			{T} 
			{\stsym} \times \ctype {T'} {\stsym}} {\ctype U 
			{\stsym}},\forall\stsym\in\stall_o$ \\
		
		\bottomrule
	\end{tabular}
	\caption{\Actor\ types.\label{tab:actor types}}
\end{table}
\Actor{} types are defined in terms of input/output signatures.
The typing of \actor s is reported in Table~\ref{tab:actor types}.
We do not show the type inference rules since they are straightforward.

From the type specification, we say each \actor{} is characterized by its 
input 
and output degrees (i.e., the cardinality of left and right-hand side of 
the 
$\to$ symbol, respectively). All \actor s but \collector\ have output degree~1, 
while \collector\ has output degree 0. All binary \actor s have input degree~2, 
\emitter\ has input degree 0 and all the other \actor s have input degree~1.

All \actor s are polymorphic with respect to \dtype{s}.
Moreover, all \actor s but \emitter\ and \collector\ are polymorphic with 
respect 
to \stype{s}.
Conversely, each \emitter\ and \collector\ \actor\ deals with one specific
\stype.%
\footnote{For example, an emitter for a finite text file would
	generate a bounded collection of strings, whereas an emitter for stream of 
	tweets
	would generate an unbounded collection of tweet objects.}

\begin{figure}
	\centering
	$\infer[\windowing]
	{\windowing op\farg \omega:\tpairio{\ctype{T}{\stsym'}}{\ctype{U}{\stsym'}}, \stsym' \in \stall_o}
	{op:\tpairio{\ctype{T}{\stsym}}{\ctype{U}{\stsym}}, \stsym \in \stall_o}$
	\caption{Unbounded extension provided by windowing\label{fig:win-typing}}
\end{figure}
As we mentioned in Section~\ref{ch:pm:coll}, a windowing \actor\ may behave as the unbounded extension of the respective bounded \actor.
This is formalized by the inference rule $\windowing$ that is reported in Figure~\ref{fig:win-typing}: given an \actor\ $op$ dealing with ordered \stype{s} (bounded or unbounded), its windowing counterpart $\windowing op$ can operate on \emph{any} ordered \stype, including \stream.
The analogous principle underlies the inference rules for all the $\windowing$ \actor s.

\subsection{\Pipe\ Types}

\begin{figure}
	%new
	\begin{subfigure}[b]{\textwidth}
		\centering
		$\infer[\pnew]{\pnew\farg op:\tau}{op:\tau}$
	\end{subfigure}
	
	%to
	\bigskip
	\begin{subfigure}[b]{\textwidth}
		\centering
		$\infer[\pto]
		{\pto\farg p\farg p_1\farg \ldots\farg p_n:\tpairio{\ctypeopt{T}{\stsym}} {\ctype V \stsym}}
		{p:\tpairio{\ctypeopt T \stsym}{\ctype U \stsym}&
			p_i:\tpairio {\ctype {U} \stsym} {(\ctypeopt {V} \stsym)_i}&
			\exists i : (\ctypeopt {V} \stsym)_i = \ctype {V} \stsym
		}$
	\end{subfigure}
	
	\bigskip
	\begin{subfigure}[b]{\textwidth}
		\centering
		$\infer[\pto_\void]
		{\pto\farg p\farg p_1\farg\ldots\farg p_n:\tpairio{\ctypeopt{T}{\stsym}} \void}
		{p:\tpairio{\ctypeopt T \stsym}{\ctype U \stsym}&
			p_i:\tpairio {\ctype {U} \stsym} \void
		}$
	\end{subfigure}
	
	%pair
	\bigskip
	\begin{subfigure}[b]{\textwidth}
		\centering
		$\infer[\ppair]
		{\ppair\farg p\farg p'\farg a:\tpairio{\ctypeopt{T}{\stsym}} {\ctypeopt V \stsym}}
		{p:\tpairio{\ctypeopt T \stsym} {\ctype U \stsym}&
			p':\tpairio \void {\ctype {U'} \stsym}&
			a:\tpairio {\ctype {U} \stsym \times \ctype {U'} \stsym} {\ctypeopt {V} 
				\stsym}
		}$
	\end{subfigure}
	
	\bigskip
	\begin{subfigure}[b]{\textwidth}
		\centering
		$\infer[\ppair']
		{\ppair\farg p\farg p'\farg a:\tpairio{\ctypeopt{T}{\stsym}} {\ctypeopt V \stsym}}
		{p:\tpairio \void {\ctype U \stsym}&
			p':\tpairio{\ctypeopt T \stsym} {\ctype {U'} \stsym}&
			a:\tpairio {\ctype {U} \stsym \times \ctype {U'} \stsym} {\ctypeopt {V} 
				\stsym}
		}$
	\end{subfigure}
	
	\bigskip
	\begin{subfigure}[b]{\textwidth}
		\centering
		$\infer[\pmerge]
		{\pmerge\farg p\farg p':\tpairio{\ctypeopt T \stsym} {\ctype U \stsym}}
		{
			{p:\tpairio{\ctypeopt T \stsym}{\ctype U \stsym}}&
			p':\tpairio \void {\ctype {U} \stsym}
		}$
	\end{subfigure}
	
	\caption{\Pipe{} typing\label{fig:pipe-typing}}
\end{figure}

\Pipe\ types are defined according to the inference rules in 
Figure~\ref{fig:pipe-typing}.
For simplicity, we use the meta-variable $\ctypeopt{T}{\stsym}$, which can be rewritten as either $\ctype{T}{\stsym}$ or~$\void$, to represent the  optional collection type\footnote{We remark the optional collection type is a mere syntactic rewriting, thus it does not represent any additional feature of the typing system.}.
The awkward rule $\pto$ covers the case in which, in a \pto\ \pipe, at least one destination \pipe\ $p_i$ has non-void output type $\ctype V \stsym$; in such case, all the destination \pipe s with non-void output type must have the same output type $\ctype V \stsym$, which is also the output type of the resulting \pipe.

%The following proposition holds for the given type system:
%\begin{prop}
%	Any \pipe{}'s input degree is equal to 0 or 1. The same holds for the 
%	output degree.
%\end{prop}
%\begin{proof}
%	By induction over \pipe{} type derivations.
%	\qedhere
%\end{proof}

Finally, we define the notion of top-level \pipe{s}, representing 
\pipe{s} that may be executed.
\begin{mydef}
	\label{def:top-level}
	A \emph{top-level} \pipe\ is a \emph{non-empty} \pipe\ of type
	$\tpairio \void \void$.
\end{mydef}

\subsection*{Running Example: Typing of \pname{word-count}}
We present the types of the \pname{word-count} components, defined in 
Section~\ref{ch:pm:design}.
We omit full type derivations since they are straightforward applications 
of 
the typing rules.

The \actor s are all unary and have the following types:
$$
\begin{array}{ll}
\aname{tokenize} &: \tpairio
{\ctype \stringtype \stsym}
{\ctype {(\stringtype \times \natsym)} \stsym},
\forall \stsym \in \stall\\

\aname{keyed-sum} &: \tpairio
{\ctype  {(\stringtype\times\natsym)} \stsym}
{\ctype  {(\stringtype\times\natsym)} \stsym},
\forall \stsym \in \stall\\

\aname{file-read} &: \tpairio
{\ctype \void \bag}
{\ctype \stringtype \bag}\\

\aname{file-write} &: \tpairio
{\ctype {(\stringtype \times \natsym)} \bag}
{\ctype \void \bag}
\end{array}
$$
\Pipe s have the following types:
$$
\begin{array}{ll}
\pname{word-count} &: \tpairio
{\ctype \stringtype \stsym}
{\ctype {(\stringtype \times \natsym)} \stsym},
\forall \stsym \in \stall\\
{\pname{file-word-count}} &: \tpairio \void \void
\end{array}
$$

We remark that \pname{word-count} is polymorphic whereas 
\pname{file-word-count} is a top-level \pipe{}.

\section{Semantics}
\label{ch:pm:semantics}

We propose an interpretation of \pipe s in terms of semantic \DF graphs, as defined in~\cite{16:bigdatasurvey:hlpp}.
Namely, we propose the following mapping:
\begin{itemize}
	\item Collections $\Rightarrow$ \DF tokens
	\item \Actor s $\Rightarrow$ \DF vertexes
	\item \pipe s $\Rightarrow$ \DF graphs
\end{itemize}
Note that collections in semantic \DF graphs are treated as a whole, thus they are mapped to single \DF tokens that flow through the graph of transformations.
In this setting, semantic \actor s (i.e., \DF vertexes) map an input collection to the respective output collection upon a single firing.

\subsection{Semantic Collections}
\label{ch:pm:semantic collections}
\DF tokens are data collections of $T$-typed elements, where $T$ is the 
\dtype 
of the collection.
Unordered collections are semantically mapped to multi-sets, whereas 
ordered collections are mapped to sequences.

We denote an unordered data collection of \dtype $T$ with the following,
``\{\;\ldots\;\}'' being interpreted as a multi-set (i.e., unordered 
collection 
with possible multiple
occurrences of elements): 
\begin{equation}
m = \Set{m_0,m_1,\ldots,m_{\card{m}-1}}
\end{equation}

A sequence (i.e., semantic ordered collection) associates a numeric \emph{timestamp} to each item, representing its temporal coordinate, in time units, with respect to time zero.
Therefore, we denote the generic item of a sequence having \dtype $T$ as
$
(t_i, s_i)
$
where $i \in \natsym$ is the position of the item in the sequence,  $t_i\in\natsym$ is the timestamp and $s_i \in T$ is the item value.
We denote an ordered data collection of \dtype $T$ with the following, where $\markeq{b}$ holds only for bounded sequences (i.e., lists):
\begin{equation}
\begin{array}{rl}
s &= \Seq{(t_0, s_0), (t_1, s_1), (t_2, s_2), \ldots\ST t_i \in \natsym, s_i \in T}\\
&= \Seq{(t_0,s_0)}\CONCAT\Seq{(t_1,s_1), (t_2,s_2), \ldots}\\
&= \Cons{(t_0,s_0)}{[(t_1,s_1), (t_2,s_2), \ldots]}\\
&\markeq{b} \Seq{(t_0,s_0),(t_1,s_1),\ldots,(t_{\card{s}-1}, s_{\card{s}-1})}
\end{array}
\end{equation}
The symbol $++$ represents the concatenation of sequence $\Seq{(t_0,s_0)}$ (head sequence)  with the sequence $\Seq{(t_1,s_1), (t_2,s_2), \ldots}$ (tail sequence).
The symbol $::$ represents the concatenation of element $(t_0,s_0)$ (head element)  with the sequence $\Seq{(t_1,s_1), (t_2,s_2), \ldots}$ (tail sequence).

We define the notion of \emph{time-ordered sequences}.
\begin{mydef}
	\label{def:time-ordered}
	A sequence $s = \Seq{(t_0, s_0), (t_1, s_1), (t_2, s_2), \ldots}$ is
	time-ordered  when the following condition is satisfied for any $i,j\in\natsym$:  $$i\leq j \Rightarrow t_i \leq t_j$$
\end{mydef}
We denote as $\ocol s$ any time-ordered permutation of $s$.
The ability of dealing with non-time-ordered sequences, which is provided by \pico, is sometimes referred as \emph{out-of-order} data processing~\cite{googlecloud:2015}.

Before proceeding to semantic \actor s and \pipe s, we define some preliminary notions about the effect of partitioning and windowing over semantic collections.

\subsubsection{Partitioned Collections}
In Section~\ref{ch:pm:actors}, we introduced partitioning policies.
In semantic terms, a partitioning policy $\pi$ defines how to group collection elements.
\begin{mydef}
	\label{def:selection}
	Given a multi-set $m$ of \dtype $T$, a function $\pi:T\to K$ and a key 
	$k \in K$, we define the $k$-selection $\select{\pi}{k}(m)$ as follows:
	\begin{equation}
	\select{\pi}{k}(m) = \{ m_i \ST x \in m_i \wedge \pi(m_i) = k \}
	\end{equation}
	Similarly, the $k$-selection $\select{\pi}{k}(s)$ of a sequence $s$ is the sub-sequence of $s$ such that the following holds:
	\begin{equation}
	\forall(t_i,s_i)\in s, (t_i, s_i)\in\select{\pi}{k}(s) \iff \pi(s_i) = k
	\end{equation}
\end{mydef}

We define the partitioned collection as the set of all groups generated according to a partitioning policy.
\begin{mydef}
	\label{def:partitioned collection}
	Given a collection $c$ and a partitioning policy $\pi$, the partitioned collection $c$ according to $\pi$, noted $c^{(\pi)}$, is defined as follows:
	\begin{equation}
	c^{(\pi)} = \Set{\select{\pi}{k}(c) \ST {k \in K}\wedge \card{\select{\pi}{k}(c)} > 0}
	\end{equation}
\end{mydef}

We remark that partitioning has no effect with respect to time-ordering.

\noindent{\bf Example:} The group-by-key decomposition, with $\pi_1$
being the left projection,\footnote{$\pi_1 (x, y) = x$} uses a special
case of selection where:
\begin{itemize}
	\item the collection has \dtype $K \times V$
	\item $\pi = \pi_1$
\end{itemize}

\subsubsection{Windowed Collections}
Before proceeding further, we provide the preliminary notion of \emph{sequence splitting}.
A splitting function $f$ defines how to split a sequence into two possibly overlapping sub-sequences, namely the \emph{head} and the \emph{tail}.
\begin{mydef}
	\label{def:collection windowing}
	Given a sequence $s$ and a splitting function $f$, the splitting of $s$ according to $f$ is: 
	\begin{equation}
	f(s) = \left(h(s), t(s)\right)
	\end{equation}
	where $h(s)$ is a bounded prefix of $s$, $t(s)$ is a proper suffix of $s$, and there is a prefix $p$ of $h(s)$ and a suffix $u$ of $t(s)$ such that $s = p\CONCAT u$.
\end{mydef}

In Section~\ref{ch:pm:windowing}, we introduced windowing policies.
In semantic terms, a windowing policy $\omega$ identifies a splitting function $f^{(\omega)}$.
Considering a split sequence $f_\omega(s)$, the head $h_\omega(s)$ represents the elements falling into the window, whereas the tail $t_\omega(s)$ represents the remainder of the sequence.

We define the windowed sequence as the result of repeated applications of windowing with time-reordering of the heads.
\begin{mydef}
	\label{def:windowed collection}
	Given a sequence $s$ and a windowing policy $w$, the windowed view of $s$ according to $w$ is:
	\begin{equation}
	s^{(\omega)} = \Seq{\ocol{s_0}, \ocol{s_1}, \ldots, \ocol{s_i}, \ldots}
	\end{equation}
	where $s_i = h_\omega(\underbrace{t_\omega(t_\omega(\ldots t_\omega}_i(s)\ldots)))$
\end{mydef}

\noindent{\bf Example:} The count-based policy $\omega = (5,2,\texttt{count})$ extracts the first 5~items from the sequence at hand and discards the first 2 items of the sequence upon sliding, whereas the tumbling policy $\omega = (5,5,\texttt{count})$ yields non-overlapping contiguous windows spanning 5 items.

%\subsubsection{Sequence Pairing}
%\begin{mydef}
%	Given two lists $u$ and $v$, we define the joined 
%	collection 
%	$\ocol s \join \ocol t$ as follows:
%	$$
%	\begin{array}{rcl}
%	\ocol s \join \ocol t &=& \Seq{\pair {s_0} {t_0}, \pair {s_0} {t_1}, 
%		\ldots, \pair {s_0} {t_{\card {\ocol t}-1}}} \CONCAT\\
%	&&\Seq{\pair {s_1} {t_0}, \pair {s_1} {t_1}, \ldots, \pair {s_1} 
%		{t_{\card 
%				{\ocol t}-1}}} \CONCAT\\
%	&&\Seq{\pair {s_{\card {\ocol s}-1}} {t_0}, \ldots, \pair {s_{\card 
%				{\ocol 
%					s}-1}} {t_{\card {\ocol t}-1}}}
%	\end{array}
%	$$
%	where $\join$ denotes the ordered join operator.
%\end{mydef}
%
%
%\begin{mydef}
%	Given two ordered collections $\ocol s$ and $\ocol t$, we define the 
%	zipped 
%	collection $\zip (\ocol s, \ocol t)$ as follows:
%	$$
%	\zip(\ocol s, \ocol t) = \Cons{\pair {s_0} {t_0} 
%	}{\zip(\Seq{s_1,\ldots}, 
%		\Seq{t_1,\ldots})}
%	$$
%	where $\zip$ denotes the zip function.
%	In case $\ocol s$ and $\ocol t$ are bounded collections, they must have 
%	the 
%	same cardinality.
%\end{mydef}
%
%The results of both joining and zipping are such that, for each resulting pair $\pair {s_i} {t_j}$, the associated timestamp is:
%$$
%\tau(\pair {s_i} {t_j}) = \text{max}(\tau(s_i),\tau(t_j))
%$$

\subsection{Semantic \Actor s}
\label{ch:pm:semantic actors}
We define the semantics of each \actor{} in terms of its behavior with respect to token processing by following the structure of Table~\ref{tab:actor types}.
We start from bounded \actor s and then we show how they can be extended to their unbounded counterparts by considering windowed streams.

\DF vertexes with one input edge and one output edge (i.e., unary \actor{}s 
with both input and output degrees equal to 1) take as input a token (i.e., 
a 
data collection), apply a transformation, and emit the resulting 
transformed 
token.
Vertexes with no input edges (i.e., \emitter)/no output edges (i.e., 
\collector) execute a routine to produce/consume an output/input token, 
respectively.
%Since we allow stateful \actor s (cf.\ Section~\ref{ch:pm:stateful actors}), 
%we 
%define vertexes to be objects rather than functions, as proposed 
%in~\cite{Lee:IEEE:P95}.

\subsubsection{Semantic Core \Actor s}
The bounded \map\ \actor\ has the following semantics:
\begin{equation}
\label{eq:strict semantic map}
\begin{array}{ll}
\map\farg f\farg m &= \Set{f(m_i)\ST m_i\in m}\\
\map\farg f\farg s &= \Seq{(t_0, f(s_0)), \ldots, (t_{\card{s}-1}, f(s_{\card{s}-1}))}
\end{array}
\end{equation}
where $m$  and $s$ are input tokens (multi-set and list, 
respectively) whereas right-hand side terms are output tokens.
In the ordered case, we refer to the above definition as \emph{strict} semantic \map, since it respects the global time-ordering of the input collection.

The bounded \flatmap\ \actor\ has the following semantics:
\begin{equation}
\begin{array}{lcl}
\flatmap\farg f\farg m &= &{\displaystyle\bigcup\Set{f(m_i)\ST m_i\in m}}\\
\flatmap\farg f\farg s &= &\Seq{(t_0, f(s_0)_0),(t_0, f(s_0)_1),\ldots,(t_0, f(s_0)_{n_0})}\CONCAT\\
&&\Seq{(t_1, f(s_1)_0),\ldots,(t_1, f(s_1)_{n_1})}\CONCAT \ldots\CONCAT \\
&&\Seq{(t_{\card{s}-1}, f(s_{\card{s}-1})_0)\ldots, (t_{\card{s}-1}, f(s_{\card{s}-1})_{n_{\card{s}-1}})}
\end{array}
\end{equation}
where $f(s_i)_j$ is the $j$-th item of the list $f(s_i)$, that is, the output of the kernel function $f$ over the input $s_i$.
Notice that the timestamp of each output item is the same as the respective input item.

The bounded \reduce\ operator has the following semantics, where $\oplus$ is both associative and commutative and, in the ordered variant, $t' = {\displaystyle\max_{(t_i,s_i) \in s}t_i}$:
\begin{equation}
\label{eq:semantic reduce}
\begin{array}{ll}
\reduce \farg \oplus \farg m &= \Set{\bigoplus\Set{m_i \in m}}\\
\reduce \farg \oplus \farg s &=\Seq{(t',(\ldots(s_0\oplus s_1)\oplus\ldots)\oplus s_{\card s - 1})}\\
&\markeq{a}\Seq{(t', s_0\oplus s_1 \oplus \ldots \oplus s_{\card s - 1})}\\
&\markeq{c}\Seq{(t', \bigoplus\Pi_2(s))}
\end{array}
\end{equation}
meaning that, in the ordered variant, the timestamp of the resulting value is the same as the input item having the maximum timestamp.
Equation~$\markeq{a}$ holds since $\oplus$ is associative and equation $\markeq{c}$ holds since it is commutative.

\GUY{In the above, when it is written $x \in s$, if timestamps are not
	visible, than it is a bit more complex than indicated, isn't it? I
	think there is also the ambiguity of what $s_i$ means: the regular
	sequence item? Or the timestamped tuple? If there was always an arrow
	above s, then this ambiguity could be avoided.}

\MD{Yes, I added the right projection to correct the last equation.
	In my intention there is no ambiguity about $s_i$, it always means the regular sequence item (without timestamp).
	Let me do a double check over the section to see if I made some misuse.}

The \aggregate\ operator has a more complex semantics, defined with respect to an \emph{arbitrary} partitioning of the input data.  Informally, given a partition $P$ of the input collection, each subset $P_i\in P$ is mapped to a local accumulator $a_i$, initialized with value $z$;
then:
\begin{enumerate}
	\item Each subset $P_i$ is folded into its local accumulator $a_i$, using $\oplus_1$;
	\item The local accumulators $a_i$ are combined using $\oplus_2$, producing a reduced value $r$;
\end{enumerate}
The formal definition---that we omit for the sake of simplicity---is similar to the semantic of \reduce, with the same distinction between ordered and unordered processing and similar considerations about associativity and commutativity of user functions.
We assume, without loss of generality, that the user parameters $z$ and $\oplus_1$ are always defined such that the resulting \aggregate{} \actor{} is partition-independent, meaning that the result is independent from the choice of the partition $P$.
%It is easy to verify this assumption is met in most use cases.

\subsubsection{Semantic Decomposition}

Given a bounded \combine{} \actor{} $op$ and a selection function $\pi:T\to K$, the partitioning \actor{} $\partitioning op$ has the following semantics over a generic collection $c$:
$$
\partitioning op \farg \pi \farg c = \Set{op \farg c' \ST c' \in c^{(\pi)}}
$$
For instance, the group-by-key processing is obtained by using the by-key 
partitioning policy (cf.\ example below definition~\ref{def:selection}).

Similarly, given a bounded \combine{} \actor{} $op$ and a windowing policy $\omega$, the windowing \actor{} $\windowing op$ has the following semantics:
\begin{equation}
\label{eq:windowing operator}
\windowing op \farg \omega \farg s = op\farg s^{(\omega)}_0 \CONCAT \ldots \CONCAT\ op\farg s^{(\omega)}_{\card{s^{(\omega)}}-1}
\end{equation}
where $s^{(\omega)}_i$ is the $i$-th list in $s^{(\omega)}$ (cf. Definition~\ref{def:windowed collection}).

As for the combination of the two partitioning mechanisms, 
\windowing\partitioning$op$, it has the following semantics:
$$
\winpar op \farg \pi\farg \omega \farg s =
\Set{\windowing op \farg \omega \farg s' \ST s' \in s^{(\pi)}}
$$
Thus, as mentioned in Section~\ref{ch:pm:actors}, partitioning first
performs the decomposition, and then processes each group on a
per-window basis.

\subsubsection{Unbounded \Actor s}
\label{ch:pm:ubops}
We remark that none of the semantic \actor s defined so far can deal with unbounded collections.
As mentioned in Section~\ref{ch:pm:actors}, we rely on windowing for extending them to the unbounded case.

Given a (bounded) windowing \combine{} \actor{} $op$, the semantics of its unbounded variant is a trivial extension of the bounded case:
\begin{equation}
\label{eq:unbounded reduce}
\windowing op \farg \omega \farg s = op\farg s^{(\omega)}_0 \CONCAT \ldots \CONCAT\ c\farg s^{(\omega)}_i \CONCAT \ldots
\end{equation}
The above incidentally also defines the semantics of unbounded windowing and partitioning \combine{} \actor s.

We rely on the analogous approach to define the semantics of unbounded \actor s in the \map{} family, but in this case the windowing policy is introduced at the semantic rather than syntactic level, since \map{} \actor s do not support decomposition.
Moreover, the windowing policy is forced to be batching (cf.\ Example below Definition~\ref{def:collection windowing}).
We illustrate this concept on \map{} \actor s, but the same holds for \flatmap{} ones. 
Given a bounded \map{} \actor, the semantics of its unbounded extension is as follows, where $\omega$ is a tumbling windowing policy:
\begin{equation}
\label{eq:weak semantic map}
\llbracket \map\farg f\farg s \rrbracket_\omega =
\map\farg f\farg s^{(\omega)}_0 \CONCAT \ldots \CONCAT\ 
\map\farg f\farg s^{(\omega)}_i \CONCAT \ldots 
\end{equation}
We refer to the above definition as \emph{weak} semantic \map (cf.\ strict semantic \map{} in Equation~\ref{eq:strict semantic map}), since the time-ordering of the input collection is partially dropped.
In the following chapters, we provide a \pico implementation based on weak semantic operators for both bounded and unbounded processing.

\subsubsection{Semantic Sources and Sinks}
Finally, \emitter/\collector\ \actor s do not have a functional semantics, 
since they produce/consume collections by interacting with the system state 
(e.g., read/write from/to a text file, read/write from/to a network socket).
From the semantic perspective, we consider each \emitter/\collector\
\actor\ as a \DF node able to produce/consume as output/input a collection of a given type, as shown in Table~\ref{tab:actor types}.
Moreover, \emitter{} \actor s of ordered type have the responsibility of tagging each emitted item with a timestamp.

\subsection{Semantic \Pipe s}
\label{ch:pm:semantic pipes}
The semantics of a \pipe{} maps it to a semantic \DF graph.
We define such mapping by induction on the \pipe{} grammar defined in Section~\ref{ch:pm:design}.
The following definitions are basically a formalization of the pictorial representation in Figure~\ref{fig:pipes}.

We also define the notion of \emph{input}, resp. \emph{output}, vertex of 
a \DF graph~$G$, denoted as $\inode G$ and $\onode G$, respectively.
Conceptually, an input node represents a \pipe{} source, whereas an output 
node represents a \pipe{} sink.

The following formalization provides the semantics of any \pico program.
\begin{itemize}
	\item ($\pnew\farg op$) is mapped to the graph $G=\pair {\Set{ op}} \emptyset$;
	moreover, one of the following three cases hold:
	\begin{itemize}
		\item $op$ is an \emitter\ \actor, then $\onode G = op$, while $\inode 
		G$ 
		is undefined
		\item $op$ is a \collector\ \actor, then $\inode G = op$, while 
		$\onode G$ 
		is undefined 
		\item $op$ is an unary \actor\ with both input and output degree 
		equal 
		to~1, then $\inode G = \onode G = op$
	\end{itemize}

	\item ($\pto\farg p\farg p_1\farg \ldots\farg p_n$) is mapped to the graph $G = \pair 
	V 
	E$ with:
	$$
	\begin{array}{ll}
	V =&{V(G_p)\cup V(G_{p_1}) \cup \ldots \cup V(G_{p_n}) \cup \Set 
		\mu}\\
	E =&E(G_p)\cup \bigcup_{i=1}^n E(G_{p_i}) \cup 
	\bigcup_{i=1}^n\Set{\pair {\onode{G_{p}}} {\inode{G_{p_i}}}} \cup\\
	&\bigcup_{i=1}^{\card {G'}}\Set{\pair {\onode{G'_{i}}} {\mu}}
	\end{array}
	$$
	where $\mu$ is a non-determinate merging node as defined in~\cite{Lee:IEEE:P95} and
	$G'=\Set{G_{p_i} \ST d_O(G_{p_i}) = 1}$;
	moreover,
	$\inode G = \inode{G_p}$ if $d_I(G_p)=1$ and undefined otherwise, while 
	$\onode{G} = \mu$ if $\card{G'} > 0$ and undefined otherwise.
	
	\item ($\ppair\farg p\farg p'\farg op$) is mapped to the graph $G = \pair V E$ with:
	$$
	\begin{array}{l}
	V = {V(G_p)\cup V(G_{p'})\cup \Set op}\\
	E = {E(G_p)\cup E(G_{p'})\cup 
		\Set{\pair{\onode{G_{p}}}{op},\pair{\onode{G_{p'}}}{op}}}
	\end{array}
	$$
	moreover, $\onode{G} = op$, while one of the following cases holds:
	\begin{itemize}
		\item $\inode G = \inode{G_{p}}$ if the input degree of $p$ is 1
		\item $\inode G = \inode{G_{p'}}$ if the input degree of $p'$ is 1
		\item $\inode G$ is undefined if both $p$ and $p'$ have output 
		degree 
		equal to 0
	\end{itemize}
	
	\item ($\pmerge\farg p\farg p'$) is mapped to the graph $G = \pair V E$ with:
	$$
	\begin{array}{l}
	V = {V(G_p)\cup V(G_{p'})\cup \Set \mu}\\
	E = {E(G_p)\cup E(G_{p'})\cup 
		\Set{\pair{\onode{G_{p}}}{\mu},\pair{\onode{G_{p'}}}{\mu}}}
	\end{array}
	$$
	where $\mu$ is a non-determinate merging node; moreover, $\onode{G} = \mu$, while one of 
	the 
	following cases holds:
	\begin{itemize}
		\item $\inode{G} = \inode{G_{p}}$ if the input degree of $p$ is 1
		\item $\inode{G} = \inode{G_{p'}}$ if the input degree of $p'$ is 1
		\item $\inode{G}$ is undefined if both $p$ and $p'$ have output 
		degree 
		equal to 0
	\end{itemize}
	
\end{itemize}

\subsection*{Running Example: Semantics of \pname{word-count}}
The tokens (i.e., data collections) flowing through the semantic \DF graph 
resulting from the \pname{word-count} \pipe\ are bags of strings 
(e.g., 
lines produced by \aname{file-read} and consumed by \aname{tokenize}) or 
bags 
of \mbox{string-$\natsym$} pairs (e.g., counts produced by \aname{tokenize} and 
consumed by \aname{keyed-sum}).
In this example, as usual, string-$\natsym$ pairs are treated as key-value 
pairs, where keys are strings (i.e., words) and values are numbers (i.e., 
counts).

By applying the semantic of \flatmap, \reduce\ and 
$\partitioning(\reduce\farg\oplus)$ to Algorithm~\ref{alg:wc model}, the result obtained is that the token being emitted by the \combine\ \actor\ is a bag of pairs $\pair w {n_w}$ for each word $w$ in the input token of the \flatmap\ \actor.

The \DF graph resulting from the semantic interpretation of the 
\pname{word-count} \pipe{} defined in Section~\ref{ch:pm:design} is $G = 
\pair 
V E$, where:
$$
\begin{array}{lll}
V&=&\Set{\text{\aname{tokenize}},\text{\aname{keyed-sum}}}\\
E&=&\Set{\pair{\text{\aname{tokenize}}}{\text{\aname{keyed-sum}}}}
\end{array}
$$
Finally, the \pname{file-word-count} \pipe{} results in the graph $G = 
\pair V 
E$ where:
$$
\begin{array}{lll}
V&=&\Set{\aname{file-read}, \aname{tokenize}, \aname{keyed-sum}, \aname{file-write}}\\
E&=&\{\pair{\aname{file-read}}{\aname{tokenize}},\\
&&\pair{\aname{tokenize}}{\aname{keyed-sum}},\\
&&\ \pair{\aname{keyed-sum}}{\aname{file-write}}\}
\end{array}
$$

\section{Programming Model Expressiveness}
\label{ch:pmexpr} % Change X to a consecutive number; for referencing this chapter elsewhere, use \ref{ChapterX}

In this section, we provide a set of use cases adapted from examples in Flink's user guide~\cite{online:flink-examples}.
Besides they are very simple examples, they exploit grouping, partitioning, windowing and \pipe s merging.
We aim to show the expressiveness of our model without using any concrete API, to demonstrate that the model is independent from its implementation.

\subsection{Use Cases: Stock Market}
The first use case is about analyzing stock market data streams.
In this use case, we:
\begin{enumerate}
	\item read and merge two stock market data streams from two sockets (algorithm~\ref{alg:stock-read})
	\item compute  statistics on this market data stream, like rolling aggregations per stock (algorithm~\ref{alg:stock-stats})
	\item emit price warning alerts when the prices change (algorithm~\ref{alg:price-warnings})
	\item compute correlations between the market data streams and a Twitter stream with stock mentions (algorithm~\ref{alg:correlate-stocks-tweets})
\end{enumerate}

\paragraph{Read from multiple sources}
\newcommand{\stocktype}{\text{StockName}}
\newcommand{\pricetype}{\text{Price}}
\begin{algorithm}[H]
	\caption{The \pname{read-price} \pipe}
	\label{alg:stock-read}
	\begin{algorithmic}[0]
		\State $\pname{read-prices} = \pnew\farg\fromsocket\farg s_1 + \pnew\farg\fromsocket\farg  s_2$
	\end{algorithmic}
\end{algorithm}
Algorithm~\ref{alg:stock-read} shows the \pname{stock-read} \pipe, which reads and merges two stock market data streams from sockets $s_1$ and $s_2$.
Assuming $\stocktype$ and $\pricetype$ are types representing stock names and prices, respectively, then the type of each \emitter\ \actor\ is the following (since \emitter\ \actor s are polymorphic with respect to \dtype):
$$
\tpairio \void {\ctype{(\stocktype \times \pricetype)}{\{\text{stream}\}}}
$$

Therefore it is also the type of \pname{read-prices} since it is a \pmerge\ of two \emitter\ \actor s of such type.

\paragraph{Statistics on market data stream}
\begin{algorithm}[H]
	\caption{The \pname{stock-stats} \pipe} % stockpricing.cpp
	\label{alg:stock-stats}
	\begin{algorithmic}[0]
		\State $\aname{min} = \reduce\farg(\lambda x y.\text{min}(x,y))$
		\State $\aname{max} = \reduce\farg(\lambda x y.\text{max}(x,y))$
		
		\State $\aname{sum-count} = \aggregate\farg$
		\parbox[t]{.6\linewidth}{%
			$(\lambda a x.((\pi_1(a)) + 1, (\pi_2(a)) + x)) \farg (0,0)$ \\
			$(\lambda a_1 a_2.(\pi_1(s_1) + \pi_1(a_2), \pi_2(a_1) + \pi_2(a_2)))$}
		
		\State $\aname{normalize} = \map\farg(\lambda x.\pi_2(x) / \pi_1(x))$
		\State $\omega = (10,5,\texttt{count})$
		\newline
		
		\State $\pname{stock-stats} = \pto\farg$
		\parbox[t]{.6\linewidth}{%
			$\pname{read-prices}$\\
			$\pnew\farg\winpar(\aname{min})\farg\pi_1\farg\omega$\\
			$\pnew\farg\winpar(\aname{max})\farg\pi_1\farg\omega$\\
			$(\pnew\farg\winpar(\aname{sum-count})\farg\pi_1\farg\omega \pseq \pnew\farg\aname{normalize})$}
	\end{algorithmic}
\end{algorithm}
Algorithm~\ref{alg:stock-stats} shows the \pname{stock-stats} \pipe, that computes three different statistics---minimum, maximum and mean---for each stock name, over the prices coming from the \pname{read-prices} \pipe.
These statistics are windowing based, since the data processed belongs to a stream possibly unbound.
The specified window policy $\omega = (10,5,\texttt{count})$ creates windows of 10 elements with sliding factor 5.

The type of \pname{stock-stats} is
$
\tpairio \void {\ctype{(\stocktype \times \pricetype)}{\{\text{stream}\}}}
$,
the same as \pname{read-prices}.

\paragraph{Generate price fluctuation warnings}
\begin{algorithm}[H]
	\caption{The \pname{price-warnings} \pipe}
	\label{alg:price-warnings}
	\begin{algorithmic}[0]
		\State $\aname{collect} = \aggregate\farg$
		\parbox[t]{.6\linewidth}{%
			$(\lambda s x.s \cup \{x\}) \farg \emptyset$\\
			$(\lambda s_1 s_2.s_1 \cup s_2)\farg$}
		
		\State $\aname{fluctuation} = \map\farg(\lambda s.\text{set-fluctuation}(s)) $
		
		\State $\aname{high-pass} = \flatmap\farg
		(\lambda \delta.\text{if }\delta \geq 0.05 \text{ then yield }\delta)$
		\State $\omega = (10,5,\texttt{count})$
		\newline
		
		\State $\pname{price-warnings} =$
		\parbox[t]{.6\linewidth}{%
			$\pname{read-prices} \pseq$\\
			$\pnew\farg\winpar(\aname{collect})\farg\pi_1\farg\omega \pseq
			\pnew\farg\aname{fluctuation}$\\
			$\pnew\farg\aname{high-pass}$
		}
	\end{algorithmic}
\end{algorithm}
Algorithm~\ref{alg:price-warnings} shows the \pipe{} \pname{price-warnings}, that generates a warning each time the stock market data within a window exhibits high price fluctuation for a certain stock name---\texttt{yield} is a host-language method that produces an element.

In the example, the \aggregate{} \actor{} \aname{fluctuation} just builds the sets, one per window, of all items falling within the window, whereas the downstream $\map$ computes the fluctuation over each set.
This is a generic pattern that allows to combine collection items by re-using available user functions defined over collective data structures.

The type of \pname{price-warnings} is again
$
\tpairio \void {\ctype{(\stocktype \times \pricetype)}{\{\text{stream}\}}}
$.

\paragraph{Correlate warnings with tweets}
\begin{algorithm}
	\caption{The \pname{correlate-stocks-tweets} \pipe}
	\label{alg:correlate-stocks-tweets}
	
	\begin{algorithmic}[0]
		\State $\pname{read-tweets} = \pnew\farg\fromtwitter \pseq \pnew\farg\aname{tokenize-tweets}$
		\State $\omega = (10,10,\texttt{count})$
		\newline
		
		\State $\pname{correlate-stocks-tweets} = \ppair$
		\parbox[t]{0.45\linewidth}{%
			$\pname{price-warnings} \farg \pname{read-tweets}$\\
			$\winpar(\aname{correlate})\farg\pi_1\farg\omega$\\
		}
	\end{algorithmic}
\end{algorithm}
Algorithm~\ref{alg:correlate-stocks-tweets} shows \pname{correlate-stocks-tweets}, a \pipe{} that generates a correlation between warning generated by \pname{price-warnings} and tweets coming from a Twitter feed.
The \pname{read-tweets} \pipe\ generates a stream of $(\stocktype \times \text{String})$ items, representing tweets each mentioning a stock name.
Stocks and tweets are paired according to a join-by-key policy (cf. definition~\ref{def:selection}), where the key is the stock name.

In the example, \aname{correlate} is a \baggregatejoin\ \actor\ that computes the correlation between two joined collections.
As we mentioned in Section~\ref{ch:pm:actors}, we rely on windowing to apply the (bounded) \baggregatejoin{} \actor\ to unbounded streams.
In the example, we use a simple tumbling policy $\omega = (10,10,\texttt{count})$ in order to correlate items from the two collections in a 10-by-10 fashion.

\section{Conclusion}
We proposed a new programming model based on \pipe s and \actor s, which are the building blocks of \pico programs, first defining the syntax of programs, then providing a formalization of the type system and semantics.

The contribution of \pico with respect to the state-of-the-art in tools for Big Data Analytics is also in the definition and formalization of a programming model that is independent from the effective API and runtime implementation.
In the state-of-the-art tools for Analytics, this aspect is typically not considered and the user is left in some cases to its own interpretation of the documentation. This happens particularly when the implementation of operators in state-of-the-art tools is conditioned in part or totally by the runtime implementation itself.

\section*{Acknowledgements}
This work was partly supported by the EU-funded project TOREADOR
(contract no.\ H2020-688797),  the EU-funded project Rephrase (contract no.\ H2020-644235),
and the 2015--2016 IBM Ph.D.\ Scholarship program. We gratefully
acknowledge Prof. Luca Padovani for his comments on the early version
of the manuscript.

%\bibliographystyle{abbrv}
%\bibliography{bib/local,bib/UniPisaTorinoGroup,bib/skeletons,bib/multicore,bib/extra,bib/mypub}

\end{document}